\documentclass[pra,aps,showpacs,superscriptaddress,onecolumn]{revtex4}
\usepackage{amsmath}
\usepackage{amstext}
\usepackage{float}
\usepackage{latexsym}
\usepackage{amssymb}
\usepackage{epsfig}

\begin{document}

\title{Multiqubit quantum phase gate using four-level superconducting
quantum interference devices coupled to superconducting resonator}
\author{Muhammad Waseem}
\affiliation{Department of Physics and Applied Mathematics, Pakistan Institute of
Engineering and Applied Sciences, Nilore, Islamabad, Pakistan}
\author{Muhammad Irfan}
\affiliation{Department of Physics and Applied Mathematics, Pakistan Institute of
Engineering and Applied Sciences, Nilore, Islamabad, Pakistan}
\author{Shahid Qamar}
\affiliation{Department of Physics and Applied Mathematics, Pakistan Institute of
Engineering and Applied Sciences, Nilore, Islamabad, Pakistan}
\date{\today}

\begin{abstract}
In this paper, we propose a scheme to realize three-qubit quantum phase gate
of one qubit simultaneously controlling two target qubits using four-level
superconducting quantum interference devices (SQUIDs) coupled to a
superconducting resonator. The two lowest levels $\left\vert 0\right\rangle $
and $\left\vert 1\right\rangle $ of each SQUID are used to represent logical
states while the higher energy levels $\left\vert 2\right\rangle $ and $%
\left\vert 3\right\rangle $ are utilized for gate realization. Our scheme
does not require adiabatic passage, second order detuning, and the
adjustment of the level spacing during gate operation which reduce the gate
time significantly. The scheme is generalized for an arbitrary n-qubit
quantum phase gate. We also apply the scheme to implement three-qubit
quantum Fourier transform.

key words: quantum phase gate, superconducting quantum interference devices
(SQUIDs), superconducting resonator, quantum Fourier transform

\pacs{85.25.Dq, 85.25.-j, 03.67.Ac}
\end{abstract}

\maketitle

\section{INTRODUCTION}

Quantum information processing has the potential ability to simulate hard
computational problems much more efficiently than classical computers. For
example factorization of large integers \cite{shore}, searching for an item
from disordered data base \cite{grover}, and phase estimation \cite{kit}.
Quantum logical gates based on unitary transformations are building blocks
of quantum computer. \textbf{The schemes for realizing two-qubit quantum
logical gates} using physical qubits such as atoms or ions in cavity QED 
\cite{scully, cirac}, superconducting devices like Josephson junctions \cite%
{blais}, cooper pair boxes \cite{mak} have been proposed. In earlier
studies, strong coupling with charge qubits \cite{ab} and flux qubits \cite%
{cpyang} was predicted in circuit QED. In some recent studies, experimental
demonstration of strong coupling in microwave cavity QED with
superconducting qubit \cite{wall, chio,paik} has been realized. The results
of these experiments make superconducting qubit cavity QED an attractive
approach for quantum information processing.

Among the superconducting state qubits, SQUIDs are promising candidate to
serve as a qubit \cite{bock}. They have long decoherence time of the order
of $1$ to $5\mu s$ \cite{hanx, vion}, design flexibility, large-scale
integration, and compatibility to conventional electronics \cite{hanx, ich,
mooj}. They can be easily embedded in the cavity while atoms or ions require
trapping techniques. Some interesting schemes for the realization of
two-qubit quantum controlled phase gates based on a cavity QED technique
with SQUIDs have been proposed \cite{yang, song, chu, w}. These studies open
a way of realizing physical quantum information processing with SQUIDs in
cavity QED.

Recently, physical realization of the multiqubit gates has gained a lot of
interest \cite{han, zubth, nori, nori2}. Algorithms for quantum computing
become complex for large qubit system. However, multiqubit quantum phase
gate reduces their complexity and can lead to faster computing. Multiqubit
quantum controlled phase gate has great importance for realizing
quantum-error- correction protocols \cite{chia}, constructing quantum
computational networks \cite{zubairy} and implementing quantum algorithms 
\cite{msz}.

In this paper, we present a scheme for the realization of three-qubit
quantum controlled phase gate of one-qubit simultaneously controlling two
qubits using four-level SQUIDs coupled to a superconducting resonator. It
may be mentioned that in an earlier study, a proposal for multiqubit phase
gate of one qubit simultaneously controlling n qubits in a cavity has been
presented \cite{nori}, which is based upon system-cavity-pulse resonant
Raman coupling, system-cavity-pulse off-resonant Raman coupling,
system-cavity off-resonant interaction and system-cavity resonant
interaction. In another study \cite{nori2}, a multiqubit phase gate based
upon the tuning of the qubit frequency or resonator frequency is proposed.
These proposals are quite general which can be applied to flux qubit systems
or SQUIDs too. The present scheme is based on system-cavity-pulse resonant
and system-cavity off-resonant interactions which can be realized using flux
qubit (SQUID) system. In this proposal, two lowest levels $%
\left|0\right\rangle$ and $\left|1\right\rangle$ of each SQUID are used to
represent the logical states while higher energy levels $\left|2\right%
\rangle $ and $\left|3\right\rangle$ are utilized for gate realization. A
single photon is created by resonant interaction of cavity field with $%
\left|2\right\rangle \leftrightarrow \left|3\right\rangle$ transition of the
control SQUID. In the presence of single photon inside the cavity,
off-resonant interaction between the cavity field and $\left|2\right\rangle
\leftrightarrow \left|3\right\rangle$ transition of each target SQUID
induces a phase shift of $e^{i\theta_n}$ to each nth target SQUID. Our
scheme has following advantages:

($1$) Controlled phase gate operation can be performed without adjusting
level spacing during gate operation, thus decoherence due to tuning of SQUID
level spacing is avoided.

($2$) The proposal does not require slowly changing Rabi frequencies (to
satisfy adiabatic passage) and use of second-order detuning (to achieve
off-resonance Raman coupling between two relevant levels), thus the gate is
significantly faster.

($3$) During the gate operation, tunneling between the levels $\left\vert
1\right\rangle $ and $\left\vert 0\right\rangle $ is not needed. The decay
of level $\left\vert 1\right\rangle $ can be made negligibly small via prior
adjustment of the potential barrier between the levels $\left\vert
1\right\rangle $ and $\left\vert 0\right\rangle $ \cite{lap}. Therefore,
each qubit can have much longer storage time.

($4$) We do not require identical coupling constants of each SQUID with the
resonator. Similarly, detuning of the cavity modes with the transition of
the relevant levels in every target SQUID is not identical, therefore, our
scheme is tolerable to inevitable non-uniformity in device parameters.

The scheme is generalized to realize n-qubit quantum controlled phase gate.
Finally, it is shown that the proposed scheme can be used to implement
three-qubit quantum Fourier transform(QFT).

\section{QUANTUM PHASE GATE}

The transformation for three-qubit quantum phase gate with one qubit
simultaneously controlling two target qubits is given by 
\begin{equation}
U_{3}\left\vert q_{1},q_{2},q_{3}\right\rangle =e^{(i{\theta _{2}}\delta
_{q_{1},1}\delta _{q_{2},1})}e^{(i{\theta _{3}}\delta _{q_{1},1}\delta
_{q_{3},1})}\left\vert q_{1},q_{2},q_{3}\right\rangle ,  \label{EQ1}
\end{equation}%
where $\left\vert q_{1}\right\rangle $, $\left\vert q_{2}\right\rangle $,
and $\left\vert q_{3}\right\rangle $ stand for basis states $\left\vert
1\right\rangle $ or $\left\vert 0\right\rangle $ for qubits 1, 2, and 3,
respectively. Here $\delta _{q_{1},1}$, $\delta _{q_{2},1}$ and $\delta
_{q_{3},1}$ are the Kronecker delta functions. It is clear from Eq. (\ref%
{EQ1}) that in three qubit quantum phase gate when control qubit $\left\vert
q_{1}\right\rangle $ is in state $\left\vert 1\right\rangle $, phase shift $%
e^{i\theta _{2}}$ induces to the state $\left\vert 1\right\rangle $ of the
target qubit $\left\vert q_{2}\right\rangle $ and phase shift $e^{i\theta
_{3}}$ to the state $\left\vert 1\right\rangle $ of target qubit $\left\vert
q_{3}\right\rangle $. When control qubit $\left\vert q_{1}\right\rangle $ is
in state $\left\vert 0\right\rangle $ nothing happens to the target qubits.
Quantum phase gate operator in Dirac notation can be written as:

\begin{eqnarray}
U_{3} &=&\left\vert 000\right\rangle \left\langle 000\right\vert +\left\vert
001\right\rangle \left\langle 001\right\vert +\left\vert 010\right\rangle
\left\langle 010\right\vert +\left\vert 011\right\rangle \left\langle
011\right\vert +\left\vert 100\right\rangle \left\langle 100\right\vert 
\notag \\
&&+e^{i\theta _{3}}\left\vert 101\right\rangle \left\langle 101\right\vert
+e^{i\theta _{2}}\left\vert 110\right\rangle \left\langle 110\right\vert
+e^{i\theta _{2}}e^{i\theta _{3}}\left\vert 111\right\rangle \left\langle
111\right\vert .
\end{eqnarray}

\begin{figure}[tbp]
\includegraphics[width=3.6 in]{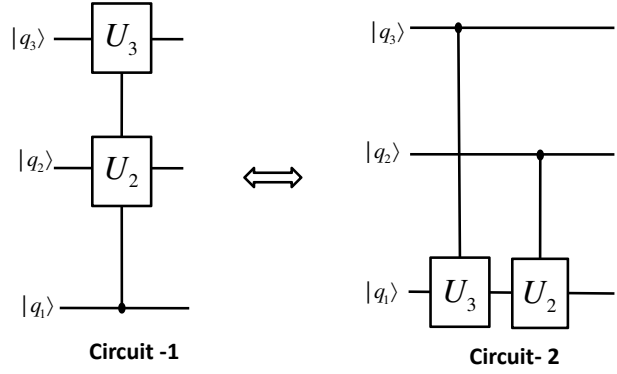}
\caption{Circuit-1 shows quantum phase gate with one qubit $\left\vert
q_{1}\right\rangle $ simultaneously controlling two target qubits $%
\left\vert q_{2}\right\rangle $ and $\left\vert q_{3}\right\rangle $.
Circuit-2 shows the two successive two-qubit controlled phase gate with
shared target qubit $\left\vert q_{1}\right\rangle $. The elements $U_{2}$
and $U_{3}$ represent two qubit controlled phase gate of phase shift $e^{i%
\protect\theta _{2}}$ and $e^{i\protect\theta _{3}}$, respectively. These
circuits are equivalent to each other.}
\label{F1}
\end{figure}

The schematic circuit diagram for quantum phase gate with one qubit
simultaneously controlling two target qubits is shown by circuit-1 in Fig. %
\ref{F1}. The circuit-2 in Fig. \ref{F1} shows the two successive two-qubit
controlled phase gate represented by $U_{2}$ and $U_{3}$ with shared target
qubit (i.e., qubit $\left\vert q_{1}\right\rangle $) but different control
qubits $\left\vert q_{2}\right\rangle $ and $\left\vert q_{3}\right\rangle $
(as shown by filled circles). The circuit-2 is known as gate decomposition
method. The elements $U_{2}$ and $U_{3}$ represent controlled phase gate
having phase shift $e^{i\theta _{2}}$ and $e^{i\theta _{3}}$, respectively.
These circuits are equivalent to each other \cite{nori} which can provide
fast implementation of QFT as discussed in Sec. $VI$.

\section{DYNAMICS OF THE SYSTEM}

Here we consider rf-SQUIDs which consists of Josephson junction enclosed by
superconducting loop. The corresponding Hamiltonian is given by \cite{rouse} 
\begin{equation}
H_{S}=\frac{Q^{2}}{2C}+\frac{(\phi -\phi _{x})^{2}}{2L}-E_{J}\cos (\frac{%
2\pi \phi }{\phi _{0}}),  \label{EQ3}
\end{equation}%
where $C$ and $L$ are junction capacitance and loop inductance,
respectively. Conjugate variables of the system are magnetic flux $\phi $
threading the ring and total charge $Q$ on capacitor. The static external
flux applied to the ring is $\phi _{x}$ and $E_{J}\equiv \frac{I_{c}\phi _{0}%
}{2\pi }$ is the Josephson coupling energy. Here $I_{c}$ is critical current
of Josephson junction and $\phi _{0}=\frac{\hbar }{2e}$ is the flux quantum.
We consider the interaction of SQUID with cavity field and microwave pulses
as discussed in the forthcoming subsections.

\subsection{Control SQUID interaction with resonator}

\begin{figure}[tbp]
\includegraphics[width=3.6 in]{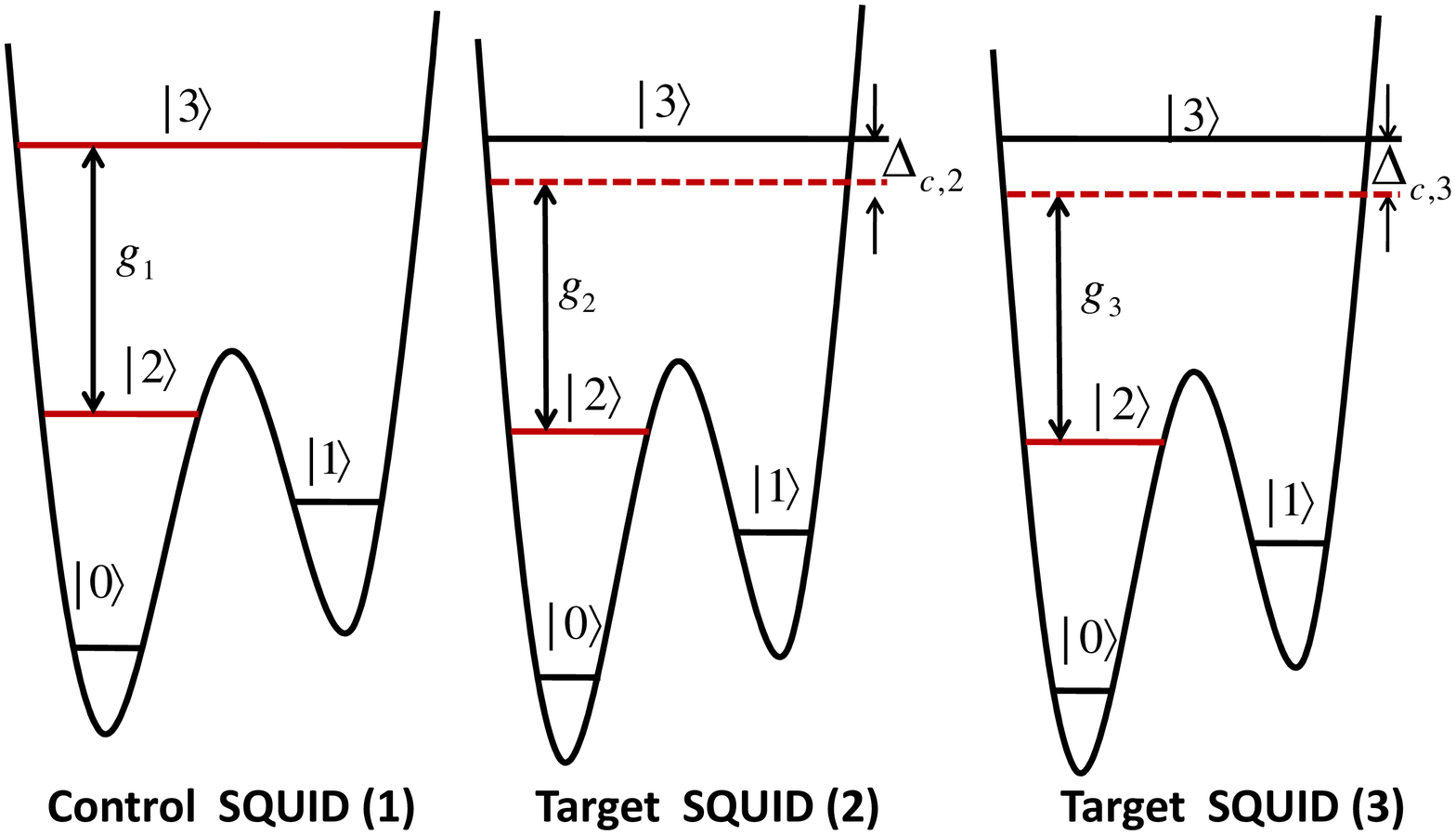}
\caption{Level diagram of control SQUID and target SQUIDs with four levels $%
\left\vert 0\right\rangle $, $\left\vert 1\right\rangle $, $\left\vert
2\right\rangle $, and $\left\vert 3\right\rangle $. The levels $\left\vert
2\right\rangle $, and $\left\vert 3\right\rangle $ of control SQUID interact
resonantly to resonator while levels $\left\vert 2\right\rangle $ and $%
\left\vert 3\right\rangle $ of each target SQUID interact off-resonantly to
the resonator. The difference between level spacing of each SQUID can be
achieved by choosing different device parameters for SQUIDs.}
\label{F2}
\end{figure}

Control SQUID is biased properly to achieve desired four-level structure by
varying the external flux \cite{w} as shown in Fig. \ref{F2}. The
single-mode of the cavity field is resonant with $\left\vert 2\right\rangle
_{1}\leftrightarrow \left\vert 3\right\rangle _{1}$ transition of control
SQUID, however, it is highly detuned from the transition between the other
levels which can be achieved by adjusting the level spacing of SQUID \cite%
{yang,lap}. Using interaction picture with rotating wave approximation one
can write the Hamiltonian of system as \cite{yang} 
\begin{equation}
H_{1}=\hbar (g_{1}a^{\dag }\left\vert 2\right\rangle _{1}\left\langle
3\right\vert +H.c),  \label{EQ4}
\end{equation}%
where $a^{\dag }$ and $a$ are photon creation and annihilation operators for
the cavity field mode of frequency $\omega _{c}$. Here $g_{1}$ is the
coupling constant between cavity field and $\left\vert 2\right\rangle
_{1}\leftrightarrow \left\vert 3\right\rangle _{1}$ transition of the
control SQUID. The evaluation of initial states $\left\vert 3\right\rangle
_{1}\left\vert 0\right\rangle _{c}$ and $\left\vert 2\right\rangle
_{1}\left\vert 1\right\rangle _{c}$ under Eq. (\ref{EQ4}) can be written as 
\begin{align}
\left\vert 3\right\rangle _{1}\left\vert 0\right\rangle _{c}& \rightarrow
cos(g_{1}t)\left\vert 3\right\rangle _{1}\left\vert 0\right\rangle
_{c}-isin(g_{1}t)\left\vert 2\right\rangle _{1}\left\vert 1\right\rangle
_{c},  \notag \\
\left\vert 2\right\rangle _{1}\left\vert 1\right\rangle _{c}& \rightarrow
cos(g_{1}t)\left\vert 2\right\rangle _{1}\left\vert 1\right\rangle
_{c}-isin(g_{1}t)\left\vert 3\right\rangle _{1}\left\vert 0\right\rangle
_{c},  \label{EQ5}
\end{align}%
where $\left\vert 0\right\rangle _{c}$ and $\left\vert 1\right\rangle _{c}$
are vacuum and single photon states of the cavity field, respectively.

\subsection{Target SQUIDs interaction with the resonator}

Suppose cavity field interacts off-resonantly with $\left\vert
2\right\rangle _{t}\leftrightarrow \left\vert 3\right\rangle _{t}$ ($%
t=2,3,...,n$) transition of each target SQUID (i.e., $\Delta _{c,t}=\omega
_{c}-\omega _{32,t}>>g_{t})$ while it is decoupled from the transition
between any other levels as shown in Fig. \ref{F2}. Here $\Delta _{c,t}$ is
the detuning between $\left\vert 2\right\rangle _{t}\leftrightarrow
\left\vert 3\right\rangle _{t}$ transition frequency $\omega _{32,t}$ of the
target SQUID ($t$) and frequency of resonator $\omega _{c}$ and $g_{t}$ is
corresponding coupling constant. The effective Hamiltonian for the system in
interaction picture can be written as \cite{guo, holland} 
\begin{equation}
H_{t}=\frac{\hbar g_{t}^{2}}{\Delta _{c,t}}(\left\vert 3\right\rangle
_{t}\left\langle 3\right\vert -\left\vert 2\right\rangle _{t}\left\langle
2\right\vert )a^{\dag }a.  \label{EQ6}
\end{equation}%
In the presence of single photon inside the cavity, the evolution of initial
states $\left\vert 2\right\rangle _{t}\left\vert 1\right\rangle _{c}$ and $%
\left\vert 3\right\rangle _{t}\left\vert 1\right\rangle _{c}$ is given by 
\begin{align}
\left\vert 2\right\rangle _{t}\left\vert 1\right\rangle _{c}& \rightarrow
e^{ig_{t}^{2}t/\Delta _{c,t}}\left\vert 2\right\rangle _{t}\left\vert
1\right\rangle _{c},  \notag \\
\left\vert 3\right\rangle _{t}\left\vert 1\right\rangle _{c}& \rightarrow
e^{-ig_{t}^{2}t/\Delta _{c,t}}\left\vert 3\right\rangle _{t}\left\vert
1\right\rangle _{c}.  \label{EQ7}
\end{align}%
It is clear that phase shift $e^{i\frac{g_{t}^{2}t}{\Delta _{c,t}}}$ and $%
e^{-i\frac{g_{t}^{2}t}{\Delta _{c,t}}}$ is induced to the state $\left\vert
2\right\rangle _{t}$ and $\left\vert 3\right\rangle _{t}$ of the target
SQUID. However, states $\left\vert 2\right\rangle _{t}\left\vert
0\right\rangle _{c}$ and $\left\vert 3\right\rangle _{t}\left\vert
0\right\rangle _{c}$ remain unchanged.

\subsection{SQUID driven by microwave pulses}

Let two levels $\left\vert i\right\rangle $ and $\left\vert j\right\rangle $
of each SQUID is driven by classical microwave pulse. The interaction
Hamiltonian in this case is \cite{yang} 
\begin{equation}
H_{\mu w}=\Omega _{ij}e^{i\varphi }\left\vert i\right\rangle \left\langle
j\right\vert +H.c,  \label{EQ8}
\end{equation}%
where $\Omega _{ij}$ is the Rabi frequency between two levels $\left\vert
i\right\rangle $ and $\left\vert j\right\rangle $ and $\varphi $ is the
phase associated with classical field. From Eq. (\ref{EQ8}) one can get the
following rotations: 
\begin{eqnarray}
\left\vert i\right\rangle &\rightarrow &cos(\Omega _{ij}t)\left\vert
i\right\rangle -ie^{-i\varphi }sin(\Omega _{ij}t)\left\vert j\right\rangle ,
\notag \\
\left\vert j\right\rangle &\rightarrow &cos(\Omega _{ij}t)\left\vert
j\right\rangle -ie^{i\varphi }sin(\Omega _{ij}t)\left\vert i\right\rangle .
\label{EQ9}
\end{eqnarray}%
In our case $\left\vert i\right\rangle \rightarrow \left\vert j\right\rangle 
$ transition corresponds to $\left\vert 0\right\rangle \rightarrow
\left\vert 2\right\rangle $, $\left\vert 1\right\rangle \rightarrow
\left\vert 2\right\rangle $ and $\left\vert 1\right\rangle \rightarrow
\left\vert 3\right\rangle $ as shown in Figs. 3-5. It may be noted that the
resonant interaction of microwave pulse with SQUID can be carried out in a
very short time by increasing the Rabi frequency of pulse i.e., intensity of
the pulse.

\section{Three-QUBIT CONTROLLED PHASE GATE}

Let us consider three-qubit controlled phase gate using three four-levels
SQUIDs coupled to a superconducting resonator. For the notation convenience,
we denote the ground level as $\left\vert 1\right\rangle \ $ and first
excited state as $\left\vert 0\right\rangle \ $ for each target SQUID as
shown in Fig. 4-5. Here we assume that the resonator mode is initially in a
vacuum state. The notation $\omega _{i,j}^{n}$ represents the microwave
frequency in resonance with the transition frequency between $\left\vert
i\right\rangle \leftrightarrow \left\vert j\right\rangle \ $ level of SQUID (%
$n=1,2,3$). Here we discuss a quantum phase gate with $\theta _{3}=\pi /4$
and $\theta _{2}=\pi /2$.

The procedure for realizing the three-qubit controlled phase gate is divided
into the following four stages of operations:

\textit{Forward resonant operation on control SQUID (1):}

\begin{figure}[tbp]
\includegraphics[width=3.6 in]{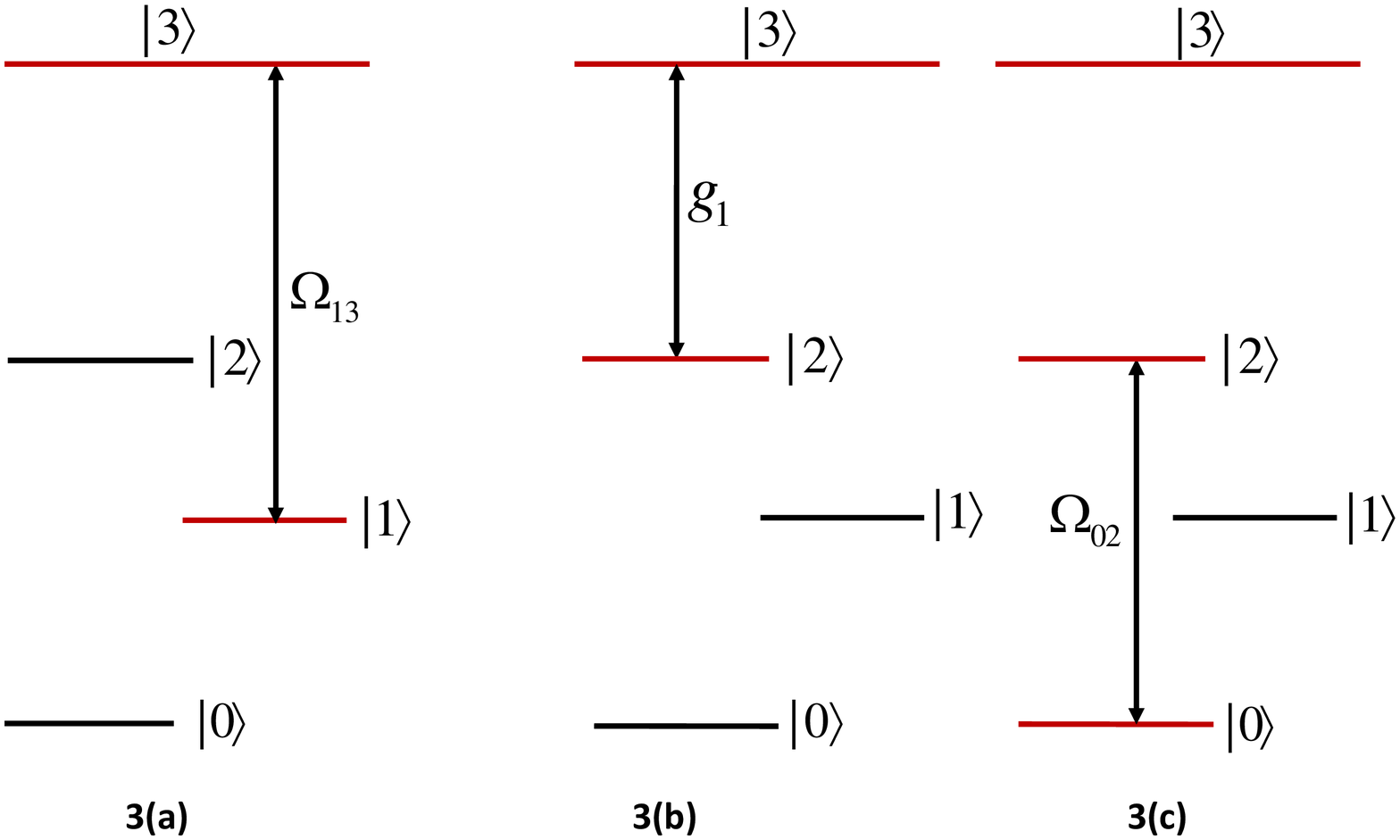}
\caption{Illustration of control SQUID(1) interacting with the resonator
mode and/or the microwave pulses during the gate operation.}
\label{F3}
\end{figure}

\textsl{Step 1:} Apply microwave pulse with frequency $f=\omega _{3,1}^{1}$
and phase $\varphi =\pi $ to the control SQUID (1). Choose time interval $%
t_{1}=\frac{\pi }{2\Omega _{13}}$ to transform the state $\left\vert
1\right\rangle _{1}$ to $i\left\vert 3\right\rangle _{1}$ as shown in Fig. %
\ref{F3}(a). The level $\left\vert 3\right\rangle _{1}$ of control SQUID (1)
is now occupied. Cavity field interacts resonantly to the $\left\vert
2\right\rangle _{1}\leftrightarrow \left\vert 3\right\rangle _{1}$
transition of control SQUID (1) shown in Fig. \ref{F3}(b). Wait for time
interval $t_{1}^{^{\prime }}=\pi /2g_{1}$ such that the transformation $%
\left\vert 3\right\rangle _{1}\left\vert 0\right\rangle _{c}\rightarrow
-i\left\vert 2\right\rangle _{1}\left\vert 1\right\rangle _{c}$ is obtained.
The overall step can be written as $\left\vert 1\right\rangle _{1}\left\vert
0\right\rangle _{c}\overset{t_{1}}{\rightarrow }i\left\vert 3\right\rangle
_{1}\left\vert 0\right\rangle _{c}\overset{t_{1}^{^{\prime }}}{\rightarrow }%
\left\vert 2\right\rangle _{1}\left\vert 1\right\rangle _{c}$. However, the
state $\left\vert 0\right\rangle _{1}\left\vert 0\right\rangle _{c}$ remains
unchanged.

\textsl{Step 2:} Apply microwave pulse with frequency $f=\omega _{2,0}^{1}$
and phase $\varphi =\pi /2$ to the control SQUID (1) as shown in Fig. \ref%
{F3}(c). We choose pulse duration $t_{2}=\pi /2\Omega _{02}$ to obtain the
transformation $\left\vert 2\right\rangle _{1}(\left\vert 0\right\rangle
_{1})\rightarrow \left\vert 0\right\rangle _{1}(-\left\vert 2\right\rangle
_{1})$. Here $\left\vert 2\right\rangle (\left\vert 0\right\rangle
)\rightarrow \left\vert 0\right\rangle (-\left\vert 2\right\rangle )$ means
the transition from $\left\vert 2\right\rangle \rightarrow \left\vert
0\right\rangle $ and $\left\vert 0\right\rangle \rightarrow -\left\vert
2\right\rangle $.

\textit{Off-resonant operation on target SQUID (2): }

\begin{figure}[tbp]
\includegraphics[width=3.5 in]{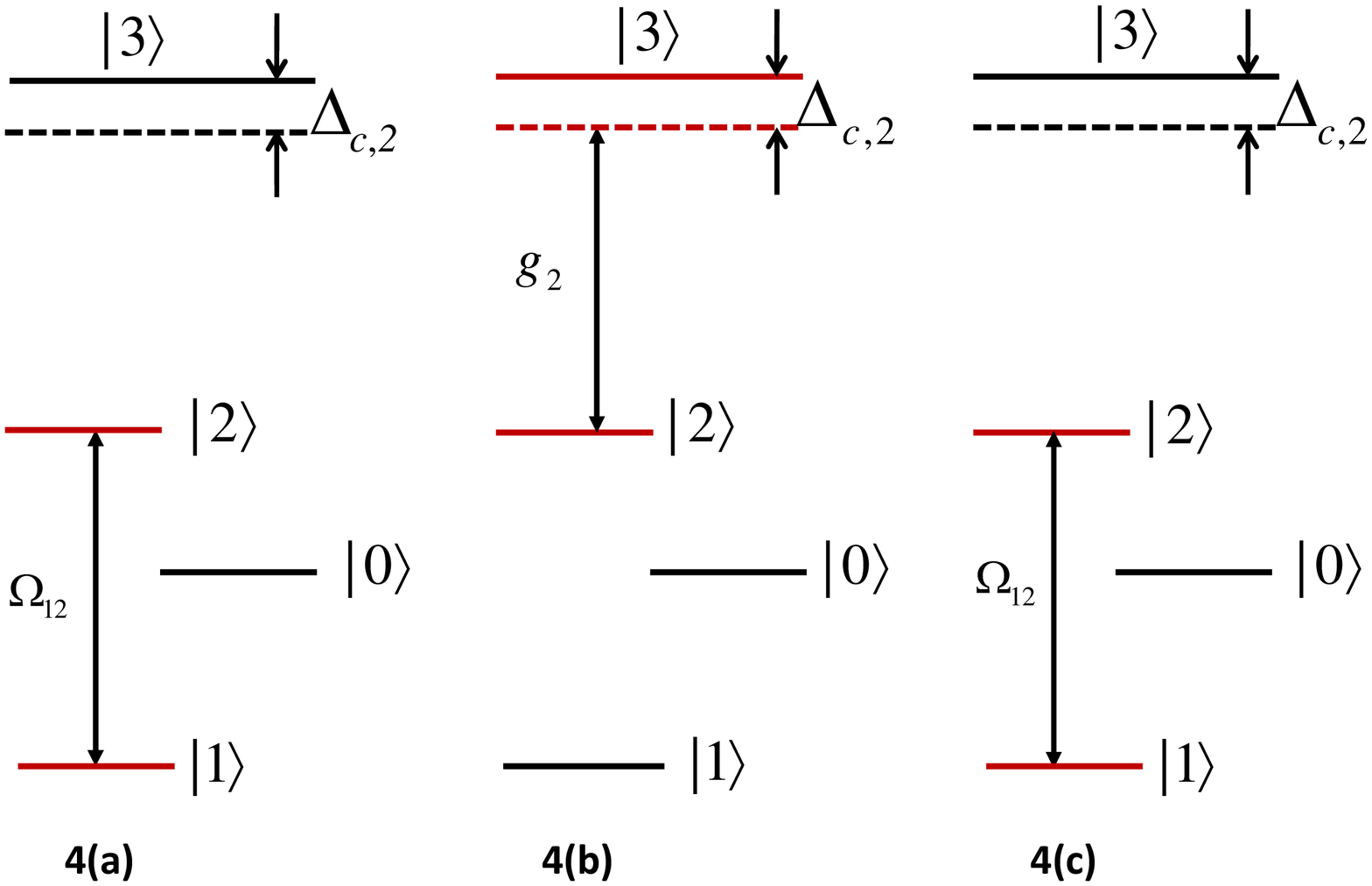}
\caption{Illustration of target SQUID(2) interacting with the resonator mode
and/or the microwave pulses during the gate operation.}
\label{F4}
\end{figure}

\textsl{Step 3:} Apply microwave pulse with frequency $f=\omega _{2,1}^{2}$
and phase $\varphi =-\pi /2$ to the SQUID (2) as shown in Fig. \ref{F4}(a).
We choose the pulse duration $t_{3}=\pi /2\Omega _{12}$ to obtain the
transformation $\left\vert 1\right\rangle _{2}(\left\vert 2\right\rangle
_{2})\rightarrow \left\vert 2\right\rangle _{2}(-\left\vert 1\right\rangle
_{2})$. It is important to mention here that, steps $2$ and $3$ can be
performed simultaneously, by setting $\Omega_{02}=\Omega _{12}$, which makes
the implementation time of both steps equal. This condition can be achieved
by adjusting the intensities of the two pulses.

\textsl{Step 4:} After the above operation, the cavity field is in a single
photon state $\left\vert 1\right\rangle _{c}$, while levels $\left\vert
2\right\rangle $ and $\left\vert 3\right\rangle $ of the control SQUID (1)
and target SQUID (3) are unpopulated. Therefore, there is no coupling of the
cavity field with control SQUID (1) and target SQUID (3). The cavity field
now interacts off-resonantly with $\left\vert 2\right\rangle _{2}\rightarrow
\left\vert 3\right\rangle _{2}$ transition of the SQUID (2) shown in Fig. %
\ref{F4}(b). It is clear from Eq. (\ref{EQ7}) that for time $t_{4}=(\pi
\Delta _{c,2})/2g_{2}^{2}$ state $\left\vert 2\right\rangle _{2}\left\vert
1\right\rangle _{c}$ evolves to $e^{i\pi /2}\left\vert 2\right\rangle
_{2}\left\vert 1\right\rangle _{c}$, where $\Delta _{c,2}$ represents the
detuning of the relevant levels of SQUID (2) with the cavity field. However,
the states $\left\vert 0\right\rangle _{2}\left\vert 0\right\rangle _{c}$, $%
\left\vert 0\right\rangle _{2}\left\vert 1\right\rangle _{c}$, and $%
\left\vert 2\right\rangle _{2}\left\vert 0\right\rangle _{c}$ remain
unchanged.

\textsl{Step 5:} Apply microwave pulse with frequency $f=\omega _{2,1}^{2}$
and phase $\varphi =\pi /2$ to the SQUID (2) as shown in Fig. \ref{F4}(c).
The transformation $\left\vert 1\right\rangle _{2}(\left\vert 2\right\rangle
_{2})\rightarrow -\left\vert 2\right\rangle _{2}(\left\vert 1\right\rangle
_{2})$ is obtained by choosing the pulse duration $t_{5}=\pi /2\Omega _{12}$.

\textit{Off-resonant operation on target SQUID (3): }

\textsl{Step 6:} Apply microwave pulse with frequency $f=\omega _{2,1}^{3}$
and phase $\varphi =-\pi /2$ to the target SQUID (3) as shown in Fig. \ref%
{F5}(a). The transformation $\left\vert 1\right\rangle _{3}(\left\vert
2\right\rangle _{3})\rightarrow \left\vert 2\right\rangle _{3}(-\left\vert
1\right\rangle _{3})$ is obtained by choosing the pulse duration $t_{6}=\pi
/2\Omega _{12}$. One can see that steps $5$ and $6$ can be performed
simultaneously, in a similar fashion as mentioned earlier.

\textsl{Step 7:} After the above operation, when the cavity field is in a
single photon state $\left\vert 1\right\rangle _{c}$, the levels $\left\vert
2\right\rangle $ and $\left\vert 3\right\rangle $ of control SQUID (1) and
target SQUID (2) are unpopulated. Under this condition, SQUIDs (1) and (2)
no longer interact with the cavity field. However, cavity field interacts
off-resonantly to $\left\vert 2\right\rangle _{3}\rightarrow \left\vert
3\right\rangle _{3}$ transition of the target SQUID (3) as shown in Fig. \ref%
{F5}(b). It is clear from Eq. (\ref{EQ7}) that for $t_{7}=(\pi \Delta
_{c,3})/4g_{3}^{2}$ state $\left\vert 2\right\rangle _{3}\left\vert
1\right\rangle _{c}$ evolves to $e^{i\pi /4}\left\vert 2\right\rangle
_{3}\left\vert 1\right\rangle _{c}$, where $\Delta _{c,3}$ represents the
detuning between the relevant levels of SQUID (3) and cavity field. However,
the states $\left\vert 0\right\rangle _{3}\left\vert 0\right\rangle _{c}$, $%
\left\vert 0\right\rangle _{3}\left\vert 1\right\rangle _{c}$, and $%
\left\vert 2\right\rangle _{3}\left\vert 0\right\rangle _{c}$ remain
unchanged.

\begin{figure}[tbp]
\includegraphics[width=3.5 in]{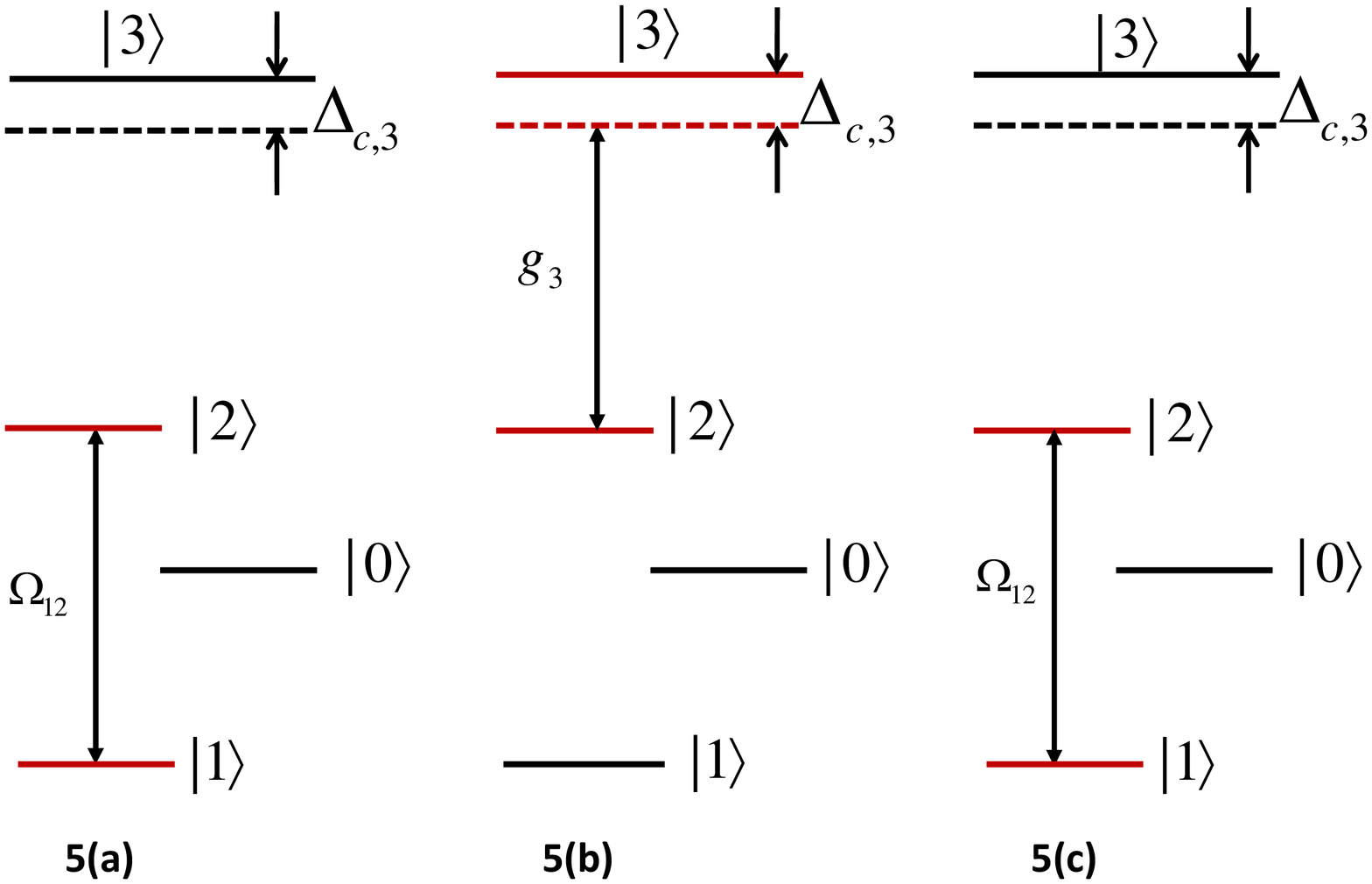}
\caption{Illustration of target SQUID(3) interacting with the resonator mode
and/or the microwave pulses during the gate operation.}
\label{F5}
\end{figure}

\textsl{Step 8:} Apply microwave pulse with frequency $f=\omega _{2,1}^{3}$
and phase $\varphi =\pi /2$ to the SQUID (3) as shown in Fig. \ref{F5}(c).
The transformation $\left\vert 1\right\rangle _{3}(\left\vert 2\right\rangle
_{3})\rightarrow -\left\vert 2\right\rangle _{3}(\left\vert 1\right\rangle
_{3})$ is obtained by choosing the pulse duration $t_{8}=\pi /2\Omega _{12}$.

\textit{Backward resonant operation on control SQUID (1):}

\textsl{Step 9:} Apply microwave pulse with frequency $f=\omega _{2,0}^{1}$
and phase $\varphi =-\pi /2$ to the control SQUID (1) shown in Fig. \ref{F3}%
(c). We choose pulse duration $t_{9}=\pi /2\Omega _{02}$ to transform state $%
\left\vert 2\right\rangle _{1}(\left\vert 0\right\rangle _{1})\rightarrow
-\left\vert 0\right\rangle _{1}(\left\vert 2\right\rangle _{1})$ for SQUID
(1). Again steps $8$ and $9$ can be performed, simultaneously.

\textsl{Step 10;} Now control SQUID (1) is in state $\left\vert
2\right\rangle $ while levels $\left\vert 2\right\rangle $ and $\left\vert
3\right\rangle $ of target SQUIDs are unpopulated when cavity field is in a
single photon state $\left\vert 1\right\rangle _{c}$. Under this condition,
both target SQUIDs no longer interact with the cavity field. Perform an
inverse operation of step(1) i.e.\textbf{,} wait for time interval $%
t_{1}^{^{\prime }}=\pi /2g_{1}$ during which resonator interacts resonantly
to the $\left\vert 2\right\rangle _{1}\leftrightarrow \left\vert
3\right\rangle _{1}$ transition of control SQUID (1) such that
transformation $\left\vert 2\right\rangle _{1}\left\vert 1\right\rangle
_{c}\rightarrow -i\left\vert 3\right\rangle _{1}\left\vert 0\right\rangle
_{c}$ is obtained (Fig. \ref{F3}(b)). Then apply microwave pulse with
frequency $f=\omega _{3,1}^{1}$ and phase $\varphi =\pi $ to the control
SQUID (1). Choose the time interval $t_{1}=\frac{\pi }{2\Omega _{13}}$ to
transform the state $\left\vert 3\right\rangle _{1}$ to $i\left\vert
1\right\rangle _{1}$ as shown in Fig. \ref{F3}(a). The over all step can be
written as $\left\vert 2\right\rangle _{1}\left\vert 1\right\rangle _{c}%
\overset{t_{1}^{^{\prime }}}{\rightarrow }-i\left\vert 3\right\rangle
_{1}\left\vert 0\right\rangle _{c}\overset{t_{1}}{\rightarrow }\left\vert
1\right\rangle _{1}\left\vert 0\right\rangle _{c}$. However, the state $%
\left\vert 0\right\rangle _{1}\left\vert 0\right\rangle _{c}$ remains
unchanged. The states of the whole system after each step of the above
mentioned operations can be summarized as

\begin{equation*}
\begin{array}{c}
\left\vert 000\right\rangle \left\vert 0\right\rangle _{c} \\ 
\left\vert 001\right\rangle \left\vert 0\right\rangle _{c} \\ 
\left\vert 010\right\rangle \left\vert 0\right\rangle _{c} \\ 
\left\vert 011\right\rangle \left\vert 0\right\rangle _{c} \\ 
\left\vert 100\right\rangle \left\vert 0\right\rangle _{c} \\ 
\left\vert 101\right\rangle \left\vert 0\right\rangle _{c} \\ 
\left\vert 110\right\rangle \left\vert 0\right\rangle _{c} \\ 
\left\vert 111\right\rangle \left\vert 0\right\rangle _{c}%
\end{array}%
\overset{1}{\rightarrow }%
\begin{array}{c}
\left\vert 000\right\rangle \left\vert 0\right\rangle _{c} \\ 
\left\vert 001\right\rangle \left\vert 0\right\rangle _{c} \\ 
\left\vert 010\right\rangle \left\vert 0\right\rangle _{c} \\ 
\left\vert 011\right\rangle \left\vert 0\right\rangle _{c} \\ 
\left\vert 200\right\rangle \left\vert 1\right\rangle _{c} \\ 
\left\vert 201\right\rangle \left\vert 1\right\rangle _{c} \\ 
\left\vert 210\right\rangle \left\vert 1\right\rangle _{c} \\ 
\left\vert 211\right\rangle \left\vert 1\right\rangle _{c}%
\end{array}%
\overset{2}{\rightarrow }%
\begin{array}{c}
-\left\vert 200\right\rangle \left\vert 0\right\rangle _{c} \\ 
-\left\vert 201\right\rangle \left\vert 0\right\rangle _{c} \\ 
-\left\vert 210\right\rangle \left\vert 0\right\rangle _{c} \\ 
-\left\vert 211\right\rangle \left\vert 0\right\rangle _{c} \\ 
\text{ \ }\left\vert 000\right\rangle \left\vert 1\right\rangle _{c} \\ 
\text{ \ }\left\vert 001\right\rangle \left\vert 1\right\rangle _{c} \\ 
\text{ \ }\left\vert 010\right\rangle \left\vert 1\right\rangle _{c} \\ 
\text{ \ }\left\vert 011\right\rangle \left\vert 1\right\rangle _{c}%
\end{array}%
\overset{3}{\rightarrow }%
\begin{array}{c}
-\left\vert 200\right\rangle \left\vert 0\right\rangle _{c} \\ 
-\left\vert 201\right\rangle \left\vert 0\right\rangle _{c} \\ 
-\left\vert 220\right\rangle \left\vert 0\right\rangle _{c} \\ 
-\left\vert 221\right\rangle \left\vert 0\right\rangle _{c} \\ 
\text{ \ }\left\vert 000\right\rangle \left\vert 1\right\rangle _{c} \\ 
\text{ \ }\left\vert 001\right\rangle \left\vert 1\right\rangle _{c} \\ 
\text{ \ }\left\vert 020\right\rangle \left\vert 1\right\rangle _{c} \\ 
\text{ \ }\left\vert 021\right\rangle \left\vert 1\right\rangle _{c}%
\end{array}%
\overset{4}{\rightarrow }%
\begin{array}{c}
\text{ \ }-\left\vert 200\right\rangle \left\vert 0\right\rangle _{c} \\ 
\text{ \ }-\left\vert 201\right\rangle \left\vert 0\right\rangle _{c} \\ 
\text{ \ }-\left\vert 220\right\rangle \left\vert 0\right\rangle _{c} \\ 
\text{ \ }-\left\vert 221\right\rangle \left\vert 0\right\rangle _{c} \\ 
\text{ \ \ \ \ \ }\left\vert 000\right\rangle \left\vert 1\right\rangle _{c}
\\ 
\text{ \ \ \ \ \ }\left\vert 001\right\rangle \left\vert 1\right\rangle _{c}
\\ 
e^{i\pi /2}\left\vert 020\right\rangle \left\vert 1\right\rangle _{c} \\ 
e^{i\pi /2}\left\vert 021\right\rangle \left\vert 1\right\rangle _{c}%
\end{array}%
\overset{5}{\rightarrow }%
\begin{array}{c}
\text{ }-\left\vert 200\right\rangle \left\vert 0\right\rangle _{c} \\ 
\text{ }-\left\vert 201\right\rangle \left\vert 0\right\rangle _{c} \\ 
\text{ \ }-\left\vert 210\right\rangle \left\vert 0\right\rangle _{c} \\ 
\text{ \ }-\left\vert 211\right\rangle \left\vert 0\right\rangle _{c} \\ 
\text{ \ \ \ \ }\left\vert 000\right\rangle \left\vert 1\right\rangle _{c}
\\ 
\text{ \ \ \ \ }\left\vert 001\right\rangle \left\vert 1\right\rangle _{c}
\\ 
e^{i\pi /2}\left\vert 010\right\rangle \left\vert 1\right\rangle _{c} \\ 
e^{i\pi /2}\left\vert 011\right\rangle \left\vert 1\right\rangle _{c}%
\end{array}%
\overset{6}{\rightarrow }%
\begin{array}{c}
\text{ }-\left\vert 200\right\rangle \left\vert 0\right\rangle _{c} \\ 
\text{ }-\left\vert 202\right\rangle \left\vert 0\right\rangle _{c} \\ 
\text{ }-\left\vert 210\right\rangle \left\vert 0\right\rangle _{c} \\ 
\text{ \ }-\left\vert 212\right\rangle \left\vert 0\right\rangle _{c} \\ 
\text{ \ \ \ \ \ }\left\vert 000\right\rangle \left\vert 1\right\rangle _{c}
\\ 
\text{ \ \ \ \ \ }\left\vert 002\right\rangle \left\vert 1\right\rangle _{c}
\\ 
e^{i\pi /2}\left\vert 010\right\rangle \left\vert 1\right\rangle _{c} \\ 
e^{i\pi /2}\left\vert 012\right\rangle \left\vert 1\right\rangle _{c}%
\end{array}%
\end{equation*}

\begin{equation}
\overset{7}{\rightarrow }%
\begin{array}{c}
\text{ \ \ \ \ \ \ \ \ }-\left\vert 200\right\rangle \left\vert
0\right\rangle _{c} \\ 
\text{ \ \ \ \ \ \ \ \ }-\left\vert 202\right\rangle \left\vert
0\right\rangle _{c} \\ 
\text{ \ \ \ \ \ \ \ \ }-\left\vert 210\right\rangle \left\vert
0\right\rangle _{c} \\ 
\text{ \ \ \ \ \ \ \ \ \ }-\left\vert 212\right\rangle \left\vert
0\right\rangle _{c} \\ 
\text{ \ \ \ \ \ \ \ \ \ \ \ \ \ }\left\vert 000\right\rangle \left\vert
1\right\rangle _{c} \\ 
\text{ \ \ \ \ \ \ }e^{i\pi /4}\left\vert 002\right\rangle \left\vert
1\right\rangle _{c} \\ 
\text{ \ \ \ \ \ \ }e^{i\pi /2}\left\vert 010\right\rangle \left\vert
1\right\rangle _{c} \\ 
e^{i\pi /4}e^{i\pi /2}\left\vert 012\right\rangle \left\vert 1\right\rangle
_{c}%
\end{array}%
\overset{8}{\rightarrow }%
\begin{array}{c}
\text{ \ \ \ \ \ \ \ \ }-\left\vert 200\right\rangle \left\vert
0\right\rangle _{c} \\ 
\text{ \ \ \ \ \ \ \ \ }-\left\vert 201\right\rangle \left\vert
0\right\rangle _{c} \\ 
\text{ \ \ \ \ \ \ \ \ }-\left\vert 210\right\rangle \left\vert
0\right\rangle _{c} \\ 
\text{ \ \ \ \ \ \ \ \ }-\left\vert 211\right\rangle \left\vert
0\right\rangle _{c} \\ 
\text{ \ \ \ \ \ \ \ \ \ \ \ \ }\left\vert 000\right\rangle \left\vert
1\right\rangle _{c} \\ 
\text{ \ \ \ \ \ \ }e^{i\pi /4}\left\vert 001\right\rangle \left\vert
1\right\rangle _{c} \\ 
\text{ \ \ \ \ \ \ }e^{i\pi /2}\left\vert 010\right\rangle \left\vert
1\right\rangle _{c} \\ 
e^{i\pi /4}e^{i\pi /2}\left\vert 011\right\rangle \left\vert 1\right\rangle
_{c}%
\end{array}%
\overset{9}{\rightarrow }%
\begin{array}{c}
\text{ \ \ \ \ \ \ \ \ \ \ }\left\vert 000\right\rangle \left\vert
0\right\rangle _{c} \\ 
\text{ \ \ \ \ \ \ \ \ \ \ \ }\left\vert 001\right\rangle \left\vert
0\right\rangle _{c} \\ 
\text{ \ \ \ \ \ \ \ \ \ \ \ \ }\left\vert 010\right\rangle \left\vert
0\right\rangle _{c} \\ 
\text{ \ \ \ \ \ \ \ \ \ \ \ \ }\left\vert 011\right\rangle \left\vert
0\right\rangle _{c} \\ 
\text{ \ \ \ \ \ \ \ \ \ \ \ \ }\left\vert 200\right\rangle \left\vert
1\right\rangle _{c} \\ 
\text{ \ \ \ \ \ \ }e^{i\pi /4}\left\vert 201\right\rangle \left\vert
1\right\rangle _{c} \\ 
\text{ \ \ \ \ \ \ }e^{i\pi /2}\left\vert 210\right\rangle \left\vert
1\right\rangle _{c} \\ 
e^{i\pi /4}e^{i\pi /2}\left\vert 211\right\rangle \left\vert 1\right\rangle
_{c}%
\end{array}%
\overset{10}{\rightarrow }%
\begin{array}{c}
\text{ \ \ \ \ \ \ \ \ \ \ }\left\vert 000\right\rangle \left\vert
0\right\rangle _{c} \\ 
\text{ \ \ \ \ \ \ \ \ \ \ \ }\left\vert 001\right\rangle \left\vert
0\right\rangle _{c} \\ 
\text{ \ \ \ \ \ \ \ \ \ \ \ }\left\vert 010\right\rangle \left\vert
0\right\rangle _{c} \\ 
\text{ \ \ \ \ \ \ \ \ \ \ \ }\left\vert 011\right\rangle \left\vert
0\right\rangle _{c} \\ 
\text{ \ \ \ \ \ \ \ \ \ \ \ \ }\left\vert 100\right\rangle \left\vert
0\right\rangle _{c} \\ 
\text{ \ \ \ \ \ \ }e^{i\pi /4}\left\vert 101\right\rangle \left\vert
0\right\rangle _{c} \\ 
\text{ \ \ \ \ \ \ }e^{i\pi /2}\left\vert 110\right\rangle \left\vert
0\right\rangle _{c} \\ 
e^{i\pi /4}e^{i\pi /2}\left\vert 111\right\rangle \left\vert 0\right\rangle
_{c}%
\end{array}
\label{eq11}
\end{equation}
Here, state $\left\vert abc\right\rangle $ is the abbreviation for states $%
\left\vert a\right\rangle _{1},\left\vert b\right\rangle _{2},\left\vert
c\right\rangle _{3}$ for SQUIDs ( 1,2, and 3 ) with a,b,c $\in \lbrack
0,1,2,3]$. It is clear from Eq. (\ref{eq11}) that three-qubit phase gate of
one qubit, simultaneously controlling two qubits can be achieved using three
SQUIDs after the above mentioned process. The total number of steps appeared
to be ten, however, they can be reduced to seven by performing some of the
steps, simultaneously as discussed earlier.

\section{GENERALIZATION TO n-QUBIT CONTROLLED PHASE GATE}

Here we discuss the generalization of our scheme to n-qubit quantum
controlled phase gate. The n-qubit quantum controlled phase gate of one
qubit simultaneously controlling n qubits is defined by the following
transformation: 
\begin{equation}
U_{n}\left\vert q_{1},q_{2},...,q_{n}\right\rangle =e^{i{\theta _{2}}\delta
_{q_{1},1}\delta _{q_{2},1}}\times e^{i{\theta _{3}}\delta _{q_{1},1}\delta
_{q_{3},1}}\times ,...,\times e^{i{\theta _{n}}\delta _{q_{1},1}\delta
_{q_{n},1}}\left\vert q_{1},q_{2},...,q_{n}\right\rangle .  \label{Q11}
\end{equation}%
Here $\delta _{q_{1},1}$, $\delta _{q_{2},1}$ , ..., $\delta _{q_{n},1}$ are
the Kronecker delta functions. First of all we define the operators
corresponding to the application of microwave pulse and cavity field
interaction to each SQUID in appropriate computational basis.

(1) Operator corresponding to the microwave pulse of phase $\pi $ applied to
control SQUID (1) in the basis $\left\vert 1\right\rangle _{1}=%
\begin{pmatrix}
0 \\ 
1%
\end{pmatrix}%
$ and $\left\vert 3\right\rangle _{1}=%
\begin{pmatrix}
1 \\ 
0%
\end{pmatrix}%
$ is $U_{\mu w}^{(1)}(\pi )=\left( 
\begin{array}{cc}
0 & i \\ 
i & 0%
\end{array}%
\right) ,$ where $\mu w$ stands for microwave.

(2) Operator corresponding to the resonant interaction of cavity field to $%
\left\vert 2\right\rangle \leftrightarrow \left\vert 3\right\rangle $
transition of control SQUID (1) in the basis $\left\vert 3\right\rangle
_{1}\left\vert 0\right\rangle _{c}=%
\begin{pmatrix}
1 \\ 
0%
\end{pmatrix}%
$ and $\left\vert 2\right\rangle _{1}\left\vert 1\right\rangle _{c}=%
\begin{pmatrix}
0 \\ 
1%
\end{pmatrix}%
$ is $U_{r}^{(1)}(\pi /2g_1)=%
\begin{pmatrix}
0 & -i \\ 
-i & 0%
\end{pmatrix}%
$, where r stands for resonance interaction.

(3) Operator corresponding to the application of microwave pulse of phase $%
-\pi /2$ for control SQUID (1) in the basis $\left\vert 2\right\rangle _{1}=%
\begin{pmatrix}
0 \\ 
1%
\end{pmatrix}%
$, $\left\vert 0\right\rangle _{1}=%
\begin{pmatrix}
1 \\ 
0%
\end{pmatrix}%
$ and for each target SQUID in the basis $\left\vert 2\right\rangle _{t}=%
\begin{pmatrix}
0 \\ 
1%
\end{pmatrix}%
$, $\left\vert 1\right\rangle _{t}=%
\begin{pmatrix}
1 \\ 
0%
\end{pmatrix}%
$ is $U_{\mu w}^{(n)}(-\pi /2)=%
\begin{pmatrix}
0 & -1 \\ 
1 & 0%
\end{pmatrix}%
$, where n=1 corresponds to the SQUID (1). In case when phase is equal to $%
\pi /2$, we have $U_{\mu w}^{(n)}(\pi /2)=(U_{\mu w}^{(n)}(-\pi /2))^{\dag }$%
.

(4) Operator corresponding to the off-resonant interaction of the cavity
field with $\left\vert 2\right\rangle \leftrightarrow \left\vert
3\right\rangle $ transition of each target SQUID in the basis \textbf{$%
\left\vert 2\right\rangle _{t}\left\vert 1\right\rangle _{c}=%
\begin{pmatrix}
0 \\ 
1%
\end{pmatrix}%
$, $\left\vert 3\right\rangle _{t}\left\vert 1\right\rangle _{c}=%
\begin{pmatrix}
1 \\ 
0%
\end{pmatrix}%
$ is $U^{(t)}(\theta _{t})=%
\begin{pmatrix}
e^{-i\theta _{t}} & 0 \\ 
0 & e^{i\theta _{t}}%
\end{pmatrix}%
$,} where\textbf{\ $\theta _{t}=\frac{g_{t}^{2}t}{\Delta _{c,t}}$.}

In order to achieve n-qubit controlled phase gate with one qubit,
simultaneously controlling n qubits, apply forward resonant operation on
control SQUID (1). As a result, single photon is created in the cavity. Then
apply off-resonant operations on target SQUID (2), target SQUID (3), up to
target SQUID (n) to induce phase shift of $e^{i\theta _{2}}$, $e^{i\theta
_{3}}$, upto $e^{i\theta _{n}}$, respectively. At the end, we apply backward
resonant operation on first control SQUID (1). Finally, single photon is
absorbed and cavity returns to its original vacuum state. Hence n-qubit
controlled phase gate can be achieved with resonator mode returning to
original vacuum state through a sequence of operations \textbf{given by} 
\begin{align}
U_{n}& =U_{\mu w}^{(1)}(\pi )\otimes U_{r}^{(1)}(\pi /2g_1)\otimes (U_{\mu
w}^{(1)}(\pi /2))^{\dag }\otimes \prod_{t=2}^{n}[U_{\mu w}^{(t)}(\pi
/2)\otimes U^{(t)}(\theta _{t})\otimes (U_{\mu w}^{(t)}(\pi /2))^{\dag }] 
\notag \\
& \otimes U_{\mu w}^{(1)}(\pi /2)\otimes U_{r}^{(1)}(\pi /2g_1)\otimes
U_{\mu w}^{(1)}(\pi ),  \label{EQ12}
\end{align}%
where $t=2,3,...,n$ stands for target SQUID and $%
\prod_{t=2}^{n}U^{(t)}=U^{(n)}\otimes ...\otimes U^{(3)}\otimes U^{(2)}$
shows off-resonant operation on each target SQUID.

The two-qubit controlled phase gate can easily be obtained from Eq. (\ref%
{EQ12}) by choosing $n=2$. In this case Eq. (\ref{EQ12}) reduces to the
following: 
\begin{align}
U_{2}& =U_{\mu w}^{(1)}(\pi )\otimes U_{r}^{(1)}(\pi /2g_{1})\otimes (U_{\mu
w}^{(1)}(\pi /2))^{\dag }\otimes U_{\mu w}^{(2)}(\pi /2)\otimes
U^{(2)}(\theta _{2})\otimes (U_{\mu w}^{(2)}(\pi /2))^{\dag }  \notag \\
& \otimes U_{\mu w}^{(1)}(\pi /2)\otimes U_{r}^{(1)}(\pi /2g_{1})\otimes
U_{\mu w}^{(1)}(\pi ).  \label{EQ13}
\end{align}%
It is clear that we need to apply forward resonant operation on control
SQUID to create single photon in the cavity. Then phase shift of $e^{i\pi }$
is induced on target SQUID by choosing interaction \textbf{time $t=\frac{\pi
\Delta _{c,2}}{g_{2}^{2}}$ }during off-resonant operation. After backward
resonant operation on control SQUID, we obtain two qubit controlled phase
gate in five steps \cite{w}. The number of steps required to implement n
qubit phase gate are $2n+1$. It may be mentioned that the number of steps
required to implement an equivalent n qubit decomposed phase gate are $5(n-1)
$.

\section{Quantum Fourier transform(QFT)}

\begin{figure}[tbp]
\includegraphics[width=3.5 in]{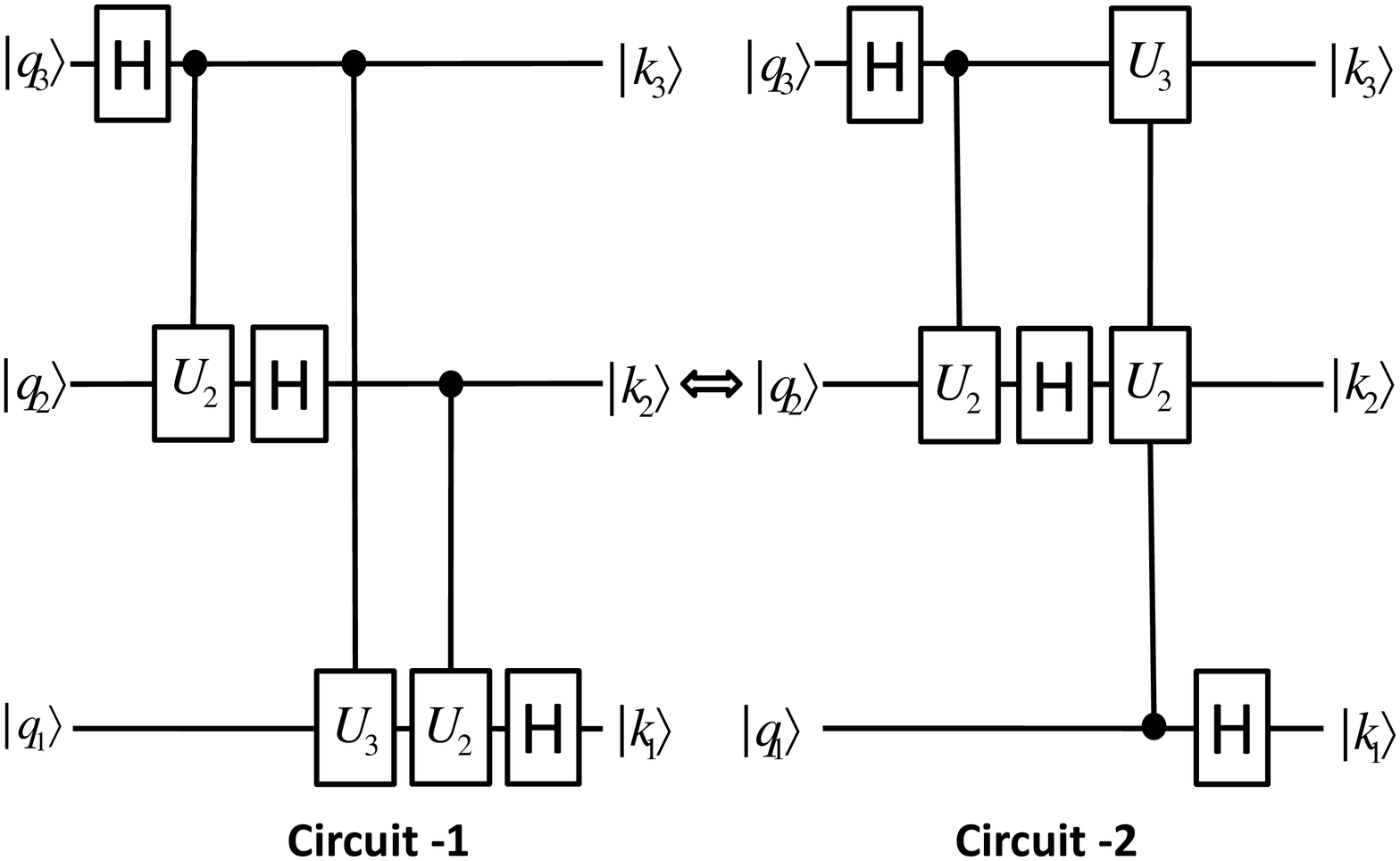}
\caption{Circuit-1 is a schematic network for three qubit QFT. The states $%
\left\vert q_{n}\right\rangle $ and $\left\vert k_{n}\right\rangle $ ($%
n=1,2,3$) represent inputs and outputs, respectively. Here H represents
Hadamard transformation and $U_{n}$ two-qubit conditional phase
transformation. The filled circles represent the control qubits. Using the
equivalent circuit shown in Fig. 1., circuit-1 reduces to circuit-2.}
\label{F6}
\end{figure}

The factorization of composite number via Shor's algorithm \cite{shore} is
an interesting example of quantum information processing. Quantum Fourier
transform lies at the heart of Shor's algorithm. Quantum Fourier transform
is a linear operator that transforms an orthogonal basis $\left\vert
k\right\rangle $ into superposition given by 
\begin{equation}
\left\vert k\right\rangle \mathbf{\rightarrow }\frac{1}{\sqrt{2^{n}}}%
\sum_{j=0}^{2^{n}-1}\mathbf{e}^{i2\pi jk2^{-n}}\left\vert j\right\rangle 
\mathbf{,}  \label{EQ14}
\end{equation}%
where n is the number of qubits. The combination of single qubit rotations
and two qubit quantum phase gate form a complex network to implement QFT for
higher qubits. For example, see circuit-1 in Fig. \ref{F6} for three qubit
QFT. However, the implementation scheme can be simplified by using
equivalent circuit as shown in Fig. \ref{F1}. For three-qubit QFT,
single-qubit Hadamard transformation, two-qubit controlled phase gate, and
three qubit quantum controlled phase gate of one qubit, simultaneously
controlling two target qubits are needed as shown by circuit-2 in Fig.\ref%
{F6}. The QFT can be accomplished in following 5 stages:

\textbf{\textit{Stage(1)}} Apply the Hadamard gate on SQUID (3). The
Hadamard rotation brings logical qubit $\left|0\right\rangle$ and $%
\left|1\right\rangle$ into superposition state i.e. $\left|0\right\rangle
\rightarrow {\frac{1}{\sqrt{2}}(\left|0\right\rangle +\left|1\right\rangle)}$
and $\left|1\right\rangle \rightarrow \frac{1}{\sqrt{2}}(\left|0\right%
\rangle -\left|1\right\rangle)$. Hadamard gate can be accomplished through
two step process that involves an auxiliary level $\left|3\right\rangle $
via method described in Ref \cite{yh}. We need two microwave pulses of
different frequencies. One is resonant to $\left|0\right\rangle%
\leftrightarrow\left|3\right\rangle$ transition frequency and other is
resonant to $\left|1\right\rangle\leftrightarrow\left|3\right\rangle$
transition frequency. Hadamard transformation can be realized through
following three steps:

\textsl{Step (a)} Apply microwave pulse with frequency $f=\omega^{3}_{1,3}$
and phase $\varphi=\pi/2 $ to SQUID (3). We choose pulse duration $t
=\pi/2\Omega_{13}$ to transform state $\left|1\right\rangle \rightarrow
-\left|3\right\rangle $ while state $\left|0\right\rangle$ remains unchanged.

\textsl{Step (b)} Apply microwave pulse with frequency $f=\omega^{3}_{0,3}$
and phase $\varphi=-\pi/2 $ to SQUID (3). We choose pulse duration $t
=\pi/4\Omega_{13}$ to obtain the transformation $\left|0\right\rangle
\rightarrow \frac{1}{\sqrt{2}}(\left|0\right\rangle +\left|3\right\rangle)$
and $\left|3\right\rangle \rightarrow \frac{1}{\sqrt{2}}(-\left|0\right%
\rangle +\left|3\right\rangle)$.

\textsl{Step (c)} Repeat the operation described in step (a) on SQUID (3) to
obtain the transformation $\left\vert 3\right\rangle \rightarrow \left\vert
1\right\rangle $ while state $\left\vert 0\right\rangle $ remains unchanged.
The above operations are summarized as 
\begin{align}
& \left\vert 0\right\rangle \overset{1}{\rightarrow }\left\vert
0\right\rangle \overset{2}{\rightarrow }\frac{1}{\sqrt{2}}(\left\vert
0\right\rangle +\left\vert 3\right\rangle )\overset{3}{\rightarrow }\frac{1}{%
\sqrt{2}}(\left\vert 0\right\rangle +\left\vert 1\right\rangle )  \notag \\
& \left\vert 1\right\rangle \overset{1}{\rightarrow }-\left\vert
3\right\rangle \overset{2}{\rightarrow }\frac{1}{\sqrt{2}}(\left\vert
0\right\rangle -\left\vert 3\right\rangle )\overset{3}{\rightarrow }\frac{1}{%
\sqrt{2}}(\left\vert 0\right\rangle -\left\vert 1\right\rangle ).
\label{EQ15}
\end{align}

\textbf{\textit{Stage (2)}} Adjust the level spacing of SQUID (3) so that
transition between levels $\left\vert 2\right\rangle _{3}$ and $\left\vert
3\right\rangle _{3}$ is resonant to the cavity field. It acts as a control
qubit. Then apply resonant forward operation on SQUID (3). A single photon
is created in the cavity which induces a phase shift of $e^{i\frac{\pi }{2}}$
on SQUID (2) through off-resonant operation.

\textbf{\textit{Stage (3)}} Repeat the same operations described in stage
(1) on SQUID (2).

\textbf{\textit{Stage (4)}} Readjust the level spacing of SQUID (3), so that
its relevant levels becomes off-resonant to the cavity field to apply
three-qubit controlled phase gate with SQUID (1) simultaneously controlling
SQUID (2) and SQUID (3) through operations described in Sec. IV.

\textbf{\textit{Stage (5)}} Repeat the same operations described in stage
(1) on SQUID (1).

Proceeding in a similar way, the proposed scheme can also be generalized up
to arbitrary n-qubit by placing n SQUIDs in a cavity. It is clear from
circuit-1 in Fig. \ref{F6} that the implementation of n-qubit QFT requires
n-Hadamard gates and $n(n-1)/2$ two-qubit phase gates. This shows the
complexity involved in implementing QFT. The situation may become much more
complicated, if complexity of calibrating and operating 4-level SQUIDs is
also taken into account. However, the complexity can be reduced using
quantum controlled phase gate of one qubit simultaneously controlling n
target qubits \cite{qftz,nori}. This is claimed only in terms of number of
steps involved and implementation time required for our proposal as compared
to the corresponding decomposed method.

It must be pointed out, that we need level adjustment for the implementation
of three-qubit QFT. However, this level adjustment is only required before
implementing stage 2 and 4 (see Fig. 6) and there is no need for level
adjustment during the implementation of phase gates. Level adjustment can be
controlled by varying external flux $\phi _{x}$ or critical current $I_{c}$ 
\cite{lap}. Thus individual SQUID can be tuned in or out of resonance with
cavity field.

\section{DISCUSSION}

The total estimated operational time for three-qubit controlled phase gate
is 
\begin{equation}
\tau =2t_{1}+2t_{1}^{^{\prime }}+t_{2}+t_{4}+t_{5}+t_{7}+t_{8}.  \label{EQ16}
\end{equation}%
On substituting the values of interaction times given in sect. IV, we
obtained 
\begin{equation}
\tau =\pi (1/\Omega _{13}+1/g_{1}+3/2\Omega _{02}+\Delta _{c,2}/2g_{2}^{2}+{%
\Delta _{c,3}}/4g_{3}^{2}).  \label{EQ17}
\end{equation}%
Here we consider without loss of generality $g_{1}\approx g_{2}\approx
g_{3}\approx 3\times 10^{9}s^{-1}$\cite{han}. On choosing $\Delta _{c,2}$%
\textbf{$=\Delta _{c,3}=10g_{3}$,} $\Omega _{02}\approx \Omega _{13}\approx
10g_{1}$, we have operational time $\tau \approx 9.16ns$. For n-qubit
controlled phase gate, we have $g_{m}=g(m=1,2,...,n)$\ and total operational
time is $\tau _{n}=\frac{\pi }{g}(\frac{22+n}{20}+\frac{10(2^{n-1}-1)}{%
2^{n-1}})$. However, if we follow the gate decomposition method then total
operation time for the equivalent circuit is given by $\tau _{n}\mathbf{=}%
\frac{\pi }{g}\mathbf{(}\frac{6(n-1)}{5}\mathbf{+}\frac{10(2^{n-1}-1)}{%
2^{n-1}}\mathbf{)}.$ The time required to implement an equivalent decomposed
three-qubit phase gate can easily be obtained from this expression, which
comes out to be $\tau \approx 10.36ns.$ In order to make a quantitative
estimate on the speed of the two approaches, next we show the plot of the
operation time as a function of number of qubits $n$ in Fig. 7. It is clear
that implementation time for the decomposed circuit increases rapidly with $n
$ as compared to the multiqubit gate. This shows that our approach is
significantly faster than performing two separate two-qubit controlled phase
gates which is quite interesting.

Here, we would like to make a comparison of our scheme with the earlier
proposal for n-target-qubit control-phase gate (NTCP) \cite{nori2}. In the
earlier study \cite{nori2}, the phase induced on each target qubit is the
same i.e., $\theta_{1}=\theta _{2}=...=\theta _{n}=\pi $, whereas in our
proposal different phases are induced on each target qubit. However, NTCP
gate can easily be realized in our scheme by using the following steps:

1. Apply step 1 of section-IV on control SQUID (1) which generates a single
photon in the cavity.

2. Apply step 2 on SQUID (1), step 3 on SQUID (2), and step 6 on SQUID (3),
simultaneously.

3. Apply off-resonant operation i.e., step 4 on SQUID (2) and step 7 on
SQUID (3), simultaneously, to obtain phase shift of $e^{i\pi }$. This can be
achieved by choosing \textbf{interaction time $t=\frac{\pi \Delta _{c,t}}{%
g_{t}^{2}}$ such }that $\frac{\Delta _{c,t}}{g_{t}^{2}}$ is the same for all
target SQUIDs.

4. Apply step 5 on SQUID (2), step 8 on SQUID (3), and step 9 on SQUID (1),
simultaneously.

5. Finally, apply step 10 on control SQUID (1), as a result single photon in
the cavity is absorbed.

Here we have considered 3 qubits, however, the scheme is applicable to
arbitrary number of qubits. Thus, NTCP gate can be realized in five steps
which is independent of the total number of qubits. The implementation time
comes out to be $12ns$.

In another study \cite{nori}, the implementation of a multiqubit phase gate
having different phases on all target qubits is proposed. The number of
steps required to implement this scheme are independent of the number of
qubits, however, the phase induced are conditional. Our proposal can also be
modified to implement a multiqubit phase gate of different phases with fixed
number of steps for arbitrary number of qubits. These modifications are
given by the following:

1. Apply step 1 on SQUID (1) as given in section-IV.

2. Apply step 2 on SQUID (1), step 3 on SQUID (2), and step 6 on SQUID (3),
simultaneously.

3. Allow both the target qubits to interact off-resonantly with the cavity
mode i.e., step 4 and 7 in section IV. The evolution is governed by Eq. (7).
In order to induce a phase of $\pi /2$ on SQUID (2) and $\pi /4$ on SQUID
(3), we have to adjust the interaction time as $\pi \Delta _{c,2}/2g_{2}^{2}$
and $\pi \Delta _{c,3}/4g_{3}^{2}$, respectively. To implement these two
steps, simultaneously, one needs both these times to be equal which can be
achieved if, $\Delta _{c,2}/g_{2}^{2}=\frac{1}{2}\Delta _{c,3}/g_{3}^{2}$.
However, this condition becomes more complex with the increase in the number
of qubits. For example, for a four-qubit gate it becomes $\Delta
_{c,2}/g_{2}^{2}=\frac{1}{2}\Delta _{c,3}/g_{3}^{2}$ and $\Delta
_{c,3}/g_{3}^{2}=\frac{1}{2}\Delta _{c,4}/g_{4}^{2}$, where $\Delta _{c,4}$
and $g_{4}$ correspond to the detuning and coupling of the fourth SQUID.

4. Apply step 5 on SQUID (2), step 8 on SQUID (3) and step 9 on SQUID (1),
simultaneously.

5. Apply step 10 on control SQUID (1), as a result single photon in the
cavity is absorbed.

It is clear that we can implement a mutiqubit phase gate having different
phases on each target qubit in five steps for arbitrary number of qubits.

Here, we discuss different issues related to the gate operations. The Level $%
\left\vert 3\right\rangle $ of control SQUID (1) is occupied in the steps 1
and 10 as discussed in section IV, which involve the microwave pulse of
frequencies $\omega _{13}$, and SQUID resonator resonant interaction. The
corresponding operational time for three-qubit controlled phase gate in step
1 or 10 is given by 
\begin{equation}
\tau _{1}=\pi (1/2\Omega _{13}+1/2g_{1}).  \label{EQ18}
\end{equation}%
It is clear that $\tau _{1}$ can be shortened sufficiently by increasing the
Rabi frequencies and coupling constant. Control SQUID (1) can be designed so
that energy relaxation time of level $\left\vert 3\right\rangle $ ($\gamma
_{3}^{-1}$) is sufficiently long as compared to the operational time. Thus
decoherence due to energy relaxation of level $\left\vert 3\right\rangle $
is negligibly small under the condition $\gamma _{3}^{-1}>>\tau _{1}$ \cite%
{w}.

The effect of dissipation during the gate operations can be neglected by
considering a high-Q resonator. The direct interaction between SQUIDs can be
negligible under the condition $H_{s,s}<< H_{s,r}H_{s,\mu\omega}$ \cite{han}%
. Here $H_{s,s}$ is the interaction energy between two nearest neighbor
SQUIDs, $H_{s,r}$ is the interaction energy between resonator and SQUID and $%
H_{s,\mu\omega}$ is the SQUID microwave interaction.

When levels $\left\vert 2\right\rangle $ and $\left\vert 3\right\rangle $
are manipulated by microwave pulses, resonant interaction as well as
off-resonant interaction between resonator mode and $\left\vert
2\right\rangle \leftrightarrow \left\vert 3\right\rangle $ transition of
each SQUID is unwanted. This effect can be minimized by setting the
condition $\Omega _{i,j}>>g_{1}$ for control SQUIDs and $\Omega
_{i,j}>>g_{t}^{2}/\Delta _{c,t}$ for target SQUIDs. The level $\left\vert
3\right\rangle $ of each target SQUID interacts off-resonantly to the cavity
field during step 4 and 7. Its occupation probability needs to be negligibly
small in order to reduce the gate error \cite{w}. The occupation probability
of level $\left\vert 3\right\rangle $ for target SQUID is given by \cite{chu}
\begin{equation}
p_{3}\approx \frac{1}{1+\frac{\Delta _{c,t}^{2}}{4g_{t}^{2}}}.  \label{EQ19}
\end{equation}%
For the choice of $\Delta _{c,t}=10g_{t}$, we have $p_{3}\approx 0.04$ which
is further reducible by increasing the ratio of $\Delta _{c,t}/g_{t}$.

The Hadamard transformation is accomplished by two microwave pulses of
different frequencies in three steps described is Sec.VI. These steps can be
implemented faster by increasing the Rabi frequency of pulses. The operation
time for Hadamard gate is around $5ns$ \cite{yh}.

\section{CONCLUSION}

In conclusion, we have presented a scheme for the realization of three-qubit
controlled phase gate with one qubit, simultaneously controlling two target
qubits using four-level SQUIDs coupled to a single-mode superconducting
microwave resonator. The scheme is based on the generation of a single
photon in the cavity mode by resonant interaction of cavity field with $%
\left\vert 2\right\rangle \leftrightarrow \left\vert 3\right\rangle $
transition of control SQUID and introduction of phase shift $e^{i\theta _{n}}
$ to each target SQUID by off-resonant interaction of the cavity field with $%
\left\vert 2\right\rangle \leftrightarrow \left\vert 3\right\rangle $
transition. Finally, backward resonant operation is applied which absorbs
the single photon as a result field inside the microwave cavity reduces to
its original vacuum state.

The proposed scheme for quantum controlled phase gate has some interesting
features, for example, it does not require adjustment of level spacing
during gate operation which reduces the cause of decoherence. The present
scheme does not require the adiabatic passage and second order detuning
which makes the gate implementation time faster. During the gate operation,
tunneling between the level $\left\vert 1\right\rangle $ and $\left\vert
0\right\rangle $ is not employed. Prior adjustment of the potential barrier
between level $\left\vert 1\right\rangle $ and $\left\vert 0\right\rangle $
can be made such that the decay of level $\left\vert 1\right\rangle $ is
negligibly small. Therefore, each qubit can have much longer storage time 
\cite{lap}. The scheme can readily be generalized to realize an arbitrary
multiqubit quantum phase gate of one qubit simultaneously controlling n
qubits. We have also applied the scheme to implement three-qubit quantum
Fourier transform.

\newpage

Caption Figure 1:

Circuit-1 shows quantum phase gate with one qubit $\left\vert
q_{1}\right\rangle $ simultaneously controlling two target qubits $%
\left\vert q_{2}\right\rangle $ and $\left\vert q_{3}\right\rangle $.
Circuit-2 shows the two successive two-qubit controlled phase gate with
shared target qubit $\left\vert q_{1}\right\rangle $. The elements $U_{2}$
and $U_{3}$ represent two qubit controlled phase gate of phase shift $%
e^{i\theta _{2}}$ and $e^{i\theta _{3}}$, respectively. These circuits are
equivalent to each other.

Caption Figure 2:

Level diagram of control SQUID and target SQUIDs with four levels $%
\left|0\right\rangle$, $\left|1\right\rangle$, $\left|2\right\rangle$, and $%
\left|3\right\rangle$. The levels $\left|2\right\rangle$, and $%
\left|3\right\rangle$ of control SQUID interact resonantly to resonator
while levels $\left|2\right\rangle$ and $\left|3\right\rangle$ of each
target SQUID interact off-resonantly to the resonator. The difference
between level spacing of each SQUID can be achieved by choosing different
device parameters for SQUIDs.

Caption Figure 3:

Illustration of control SQUID(1) interacting with the resonator mode and/or
the microwave pulses during the gate operation.

Caption Figure 4:

Illustration of target SQUID(2) interacting with the resonator mode and/or
the microwave pulses during the gate operation.

Caption Figure 5:

Illustration of target SQUID(3) interacting with the resonator mode and/or
the microwave pulses during the gate operation.

Caption Figure 6:

Circuit-1 is a schematic network for three qubit QFT. The states $\left\vert
q_{n}\right\rangle $ and $\left\vert k_{n}\right\rangle $ ($n=1,2,3$)
represent inputs and outputs, respectively. Here $H$ represents Hadamard
transformation and $U_{n}$ two-qubit conditional phase transformation. The
filled circles represent the control qubits. Using the equivalent circuit
shown in Fig. 1., circuit-1 reduces to circuit-2.

Caption Figure 7: Plot of the gate implementation time against the number of
qubits.

\newpage

\begin{figure}[tbp]
\includegraphics[width=3.8 in]{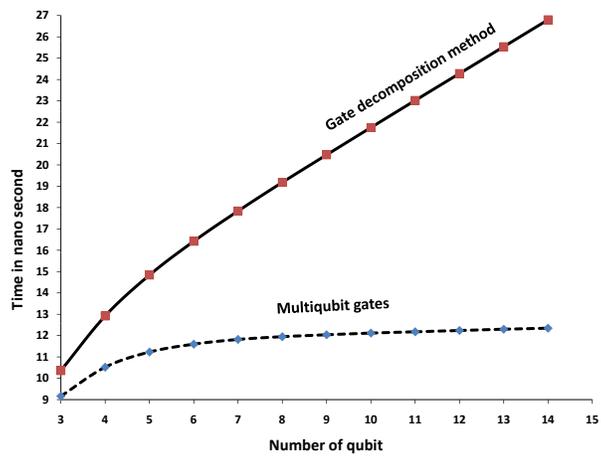}
\caption{Plot of the gate implementation time against the number of qubits.}
\label{F7}
\end{figure}

\end{document}